\documentclass[preprint]{elsarticle}
\usepackage{setspace}
\usepackage{graphicx}
\usepackage{natbib}
\usepackage{amsmath, amssymb}

\begin{document}

\begin{frontmatter}
\title{Chaotic dust dynamics and implications for the hemispherical color asymmetries of the Uranian satellites}
\author[label1]{Daniel Tamayo\corref{cor1}}
\ead{dtamayo@astro.cornell.edu}
\cortext[cor1]{Corresponding Author}
\address[label1]{Department of Astronomy, Cornell University, Ithaca, NY 14853, USA}

\author[label1,label2]{Joseph A. Burns}
\address[label2]{School of Mechanical Engineering, Cornell University, Ithaca, NY 14853, USA}
\address[label3]{Department of Astronomy, University of Maryland, College Park, MD 20742, USA}
\author[label3]{Douglas P. Hamilton}

\begin{abstract}
Dust grains generated by the Uranian irregular satellites will undergo chaotic large-amplitude eccentricity oscillations under the simultaneous action of radiation forces and the highly misaligned quadrupole potentials of the oblate planet and distant Sun.  From a suite of orbital histories, we estimate collision probabilities of dust particles with the regular satellites and argue that this process may explain the observed hemispherical color asymmetries of the outermost four regular satellites of Uranus.
\end{abstract}

\end{frontmatter}

\doublespacing 

\section{Introduction}
\cite{Buratti91} found that all five of the primary Uranian satellites, excluding the innermost Miranda, exhibit systematic leading-trailing color asymmetries of roughly 2-23\% that increase with distance from the primary.  In particular, the leading hemispheres of these tidally locked satellites (pointing in the direction of motion) are redder than their respective trailing hemispheres.  Explaining the origin of this phenomenon is challenging.  We briefly consider several exogenic possibilities, dismissing endogenic alternatives due to the difficulty of geological processes producing an effect coinciding with the satellite's apex of motion.  Exogenic hypotheses can be divided into two categories:  sources from within the Uranian system, and sources from beyond.

For explanations that rely on particles that come from beyond the Uranian system, we consider alteration by:  (i) interplanetary dust particles (IDPs) \citep[see][for an analysis of the Saturnian system]{Cook70}, (ii) interstellar dust particles (ISDPs) \citep{Landgraf00}, (iii) solar radiation \citep{Hodyss09}, and (iv) cosmic rays \citep{Johnson90}.  A tidally locked satellite on a circular orbit (like the primary Uranian satellites, to an excellent approximation) rotates on its axis once per orbit at a constant rate.  For (iii) and (iv), the incoming particle velocity is large enough that the satellite's motion is negligible.  Furthermore, the paths of these satellites over an orbital period ($\lesssim$ 2 weeks) are small compared to the distances over which the incoming particles travel.  If one therefore ignores the moon's orbital motion and simply imagines it rotating in place, one can see that the leading and trailing sides will spend an equal amount of time facing each inertial direction.  Thus, a distribution of fast incoming particles like solar photons and cosmic rays cannot generate a hemispherical leading/trailing asymmetry.  

On the other hand, if the incoming particles have speeds comparable to a satellite's orbital speed ($\sim 5$ km/s), the moon's motion introduces a detectable asymmetry.  Like a car's windshield, the satellite's leading side will accumulate more material relative to its trailing side the faster it ploughs through the rain of particles.  This is certainly the case for IDPs, whose speeds relative to the satellites is $v_{Rel} \sim 10$ km/s, and less so for ISDPs ($v_{Rel} \sim 30$ km/s).  However, this argument predicts that inner satellites with faster orbital velocities should exhibit stronger asymmetries, which is opposite to the trend found by \cite{Buratti91}.   Furthermore, gravitational focusing by Uranus would intensify the flux of IDPs and ISDPs nearer to the planet, which again runs counter to the observed pattern.  Thus, (iii) and (iv) cannot produce leading/trailing asymmetries, and while (i) and (ii) in principle could, they would generate the opposite trend with semimajor axis to what is observed.  It therefore seems unlikely that the source of the Uranian regular satellites' leading/trailing asymmetries lies beyond the Uranian system.

Explanations from within the Uranian system include magnetospheric effects \citep[see][for a discussion in the Saturnian system]{Schenk11}, and the infall of dust that originates at the irregular satellites.  Though resonant phenomena could in principle complicate the picture, magnetospheric effects would generally incorrectly predict a larger effect on inner moons, since all the satellites are beyond the co-rotation radius and the magnetic field strength falls off rapidly with semimajor axis \citep{Buratti91}.  We are therefore left with the last hypothesis, dust from the irregular satellites, which we evaluate in the remainder of this paper.  In fact, though the currently known Uranian irregular moons lay undiscovered at the time, \cite{Buratti91} argued that such dust from unseen outer satellites could account for the observed hemispherical differences.  

\subsection{The Irregular Satellites}

To date, nine irregular satellites have been found around Uranus, and many more around the other giant planets.  In contrast to the large regular satellites nestled close to their planets, the irregular satellites are a separate population of distant, small moons.  These bodies, rather than forming in a circumplanetary disk, are thought to have been captured by their respective planets' gravity (perhaps with the aid of drag forces) early in the Solar System's history \cite[see][and references therein]{Nicholson08}.  As a result, the irregulars' orbits form a distant swarm of mutually inclined, highly elliptical, crossing orbits. This  suggests an intense collisional history that would have generated much debris and dust, particularly at early times \citep{Bottke10}.  Furthermore, micrometeoroid bombardment of the irregular satellite surfaces would contribute further dust over the age of the Solar System.

\begin{figure}[!ht]
\includegraphics[width=12cm]{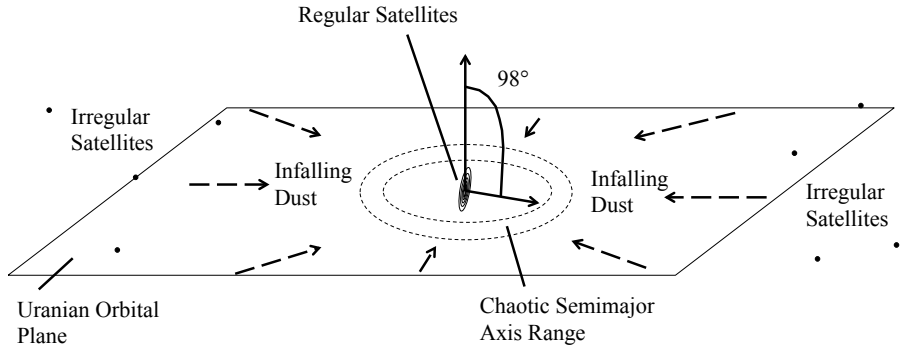}
\caption{\label{cartoon}  Schematic diagram showing the geometry of the Uranian system.  The regular satellite orbits lie in the planet's equatorial plane, which is inclined by $98^{\circ}$ to the planet's orbital plane, shown by the surrounding rectangle.  The irregular satellites (black dots) lie at large distance from the planet on inclined orbits to the planet's orbital plane.  They have only been drawn on the left and right for clarity, but there would also be moons at phases in their orbit such that they lie at the bottom and top edges of the plane pictured.  Dust from these satellites will spiral inward over millions of years through P-R drag, eventually entering a chaotic semimajor axis range schematically depicted by two concentric, dashed rings.  Upon doing so, the dust orbits will undergo chaotic large-amplitude oscillations in eccentricity and inclination.  See Sec. \ref{Dynamics} for details.}
\end{figure}

Dust particles of radius $10\mu$m will then slowly migrate inward through Poynting-Robertson (P-R) drag on a timescale of five million years, with the timescale for larger particles scaling linearly with grain radius \citep{Burns79}.  Upon reaching the inner Uranian system, this dust will coat the regular satellites.  We evaluate the dust grains' ability to generate leading/trailing asymmetries below.  Figure \ref{cartoon} shows this process schematically.  Note that the irregular satellites, which are dominantly affected by solar perturbations, lie (very roughly) symmetrically about the planet's orbital plane, while the regular satellites lie in the planet's equatorial plane, which, due to Uranus' extreme obliquity, is $98^{\circ}$ away!  The diagram also displays a chaotic range in semimajor axis that is more fully described below.

We pause to caution that several important aspects of these processes are poorly constrained.  For example, the lifetimes of dust particles orbiting planets in the outer Solar System are very uncertain.  The main mechanisms for grain destruction are sputtering, shattering by micrometeoroids, and sublimation.  The last of these is not thought to be important at Uranus and Neptune, and \cite{Burns01} give sputtering and shattering timescales of $\sim 10^{5\pm2}$ and $\sim 10^{6\pm2}$ years for 1-micron particles orbiting Uranus in its magnetosphere, respectively.  However, dust from the irregular satellites lives out its life in a very different environment to typically-considered circumplanetary grains.  Because the irregular satellites ($\sim 200-800$ Uranian radii, $R_p$) reside far beyond the magnetopause ($\sim 20 R_p$), sputtering should be much less important.  Also, far out in the Uranian gravity well, IDPs and ISDPs are not as gravitationally focused and have reduced orbital speeds, resulting in longer collision timescales.  However, depending on the optical depth of the generated dust cloud, one may have to consider mutual collisions between grains \citep{Tamayo11}.  Given our crude knowledge, it is not clear whether lifetimes of such particles can be very long ($\sim 100$ Myr), or whether large particles will be eroded into smaller particles that evolve inward faster and have longer collisional lifetimes \citep{Burns01}.

A second uncertainty is the total supply of dust available from the irregular satellites.  If this quantity is much smaller than the mass of IDPs striking the regular satellites, it would seem dubious to suppose that the irregulars could be responsible for the color asymmetries.  \cite{Cuzzi98} estimate an IDP mass flux $\sim 5 \times 10^{-16}$ kg m$^{-2}$s$^{-1}$ of IDPs in the outer Solar System.  This corresponds to $\sim 10^{14}$ kg on each of the regular satellites over 5 Gyr.  As for the irregulars, \cite{Bottke10} estimate that, over the Solar System's history, these satellites would produce $\sim 10^{20}$ kg of dust solely through mutual collisions (this number would be enhanced by micrometeoroid bombardment).  This value, however, is very uncertain since it assumes a number and distribution of primordial irregular satellites that is poorly constrained (it draws initial conditions from models of irregular satellite capture during a Nice-model reshuffling of the planets).  Furthermore, as discussed above, it is then not clear what fraction of this dust will survive on its way inward.  Nevertheless, we find below that the vast majority of surviving grains will strike one of the regular satellites; it therefore seems plausible that irregular satellite debris represents the dominant source of micrometeoroids impacting the regular satellites.

Finally, the precise mechanism through which incoming dust particles alters the satellites' surface color is unclear.  Does the altered color represent a contribution from the dust material?  Is the satellite's regolith mineralogy altered by the micrometeoroid impacts due to vaporization and/or melting?  Or is it something else?  Presumably the answer involves all three.

In the end, we take the view that the above considerations are unfortunately too uncertain to be of much guidance.  Yet if, as we have argued, other sources are unable to account for the hemispheric asymmetries and if, as we will try to show, irregular satellite dust can, then perhaps this provides constraints on these quantities that are so difficult to estimate, like the grain lifetimes and the total dust mass generated by the irregulars.  

\subsection{Dynamics} \label{Dynamics}
\cite{Buratti91} favored an infalling-dust explanation of the hemispherical asymmetries by analogy to arguments for the Saturnian satellite Iapetus \citep{Soter74, Tosi10, Tamayo11}, which seems to display a much more extravagant hemispherical albedo pattern as a result of subsequent runaway water ice sublimation \citep{Spencer10}.  This analogy has been strengthened by the recent discovery on Iapetus of a color dichotomy similar to those observed on the Uranian satellites \cite{Denk10}.  However, a closer look at investigations in the Saturnian system reveals an important difficulty.

\cite{Tosi10} and \cite{Tamayo11} find that Iapetus, the outermost Saturnian regular satellite, intercepts the vast majority of dust grains from the irregular satellites.  This should also be the case in the Uranian system.  Because the collision time with each satellite ($\tau_{Col} \sim 10^5 $yrs) is much shorter than the P-R decay timescale on which the dust moves inward ($\tau_{P-R} \sim 5 \times 10^{6}$ yrs for the smallest particles), one would expect almost no dust to penetrate inside the orbit of the outermost moon Oberon---yet three moons further in are observed to also exhibit leading/trailing asymmetries.  The key breakdown in the analogy between the two planetary systems is a dynamical instability in a chaotic range of semimajor axes that occurs owing to Uranus' extreme obliquity of $98^{\circ}$.  The goal of this paper is to investigate the effects of this instability to assess whether it is consistent with the observed hemispherical color asymmetries.

\cite{Tremaine09} study the dynamics of circumplanetary-particle orbits under the combined effect of the quadrupole potentials due to the planet's oblateness and the solar tide.  They find that around planets with obliquities exceeding $71.072^{\circ}$, particles with orbits starting far from the primary in the planet's orbital plane that are slowly brought inward become unstable over a range of semimajor axes.  In this unstable region, orbits undergo chaotic, large-amplitude oscillations in eccentricity and inclination on a secular timescale that is $\tau_{Sec} \sim 10^4$ yrs for circumuranian particles.  The unstable range roughly coincides with the location where the strengths of the two perturbations are comparable, which around Uranus, is $\sim 75 R_p$ \citep{Tamayo13}.

This provides a possible mechanism for creating the color dichotomies observed on the four outermost Uranian satellites.  Instead of slowly drifting past Oberon through P-R drag, the grains' pericenters abruptly plunge inward upon entering the unstable semimajor axis range.  This would (nearly simultaneously) spread dust across all the inner moons rather than predominantly concentrating it on the outermost satellite.  However, the dynamical results of \cite{Tremaine09} cannot be immediately applied to dust particles, as dust grains are also strongly perturbed by radiation pressure \citep{Burns79}. 

\begin{figure}[!ht]
\includegraphics[width=12cm]{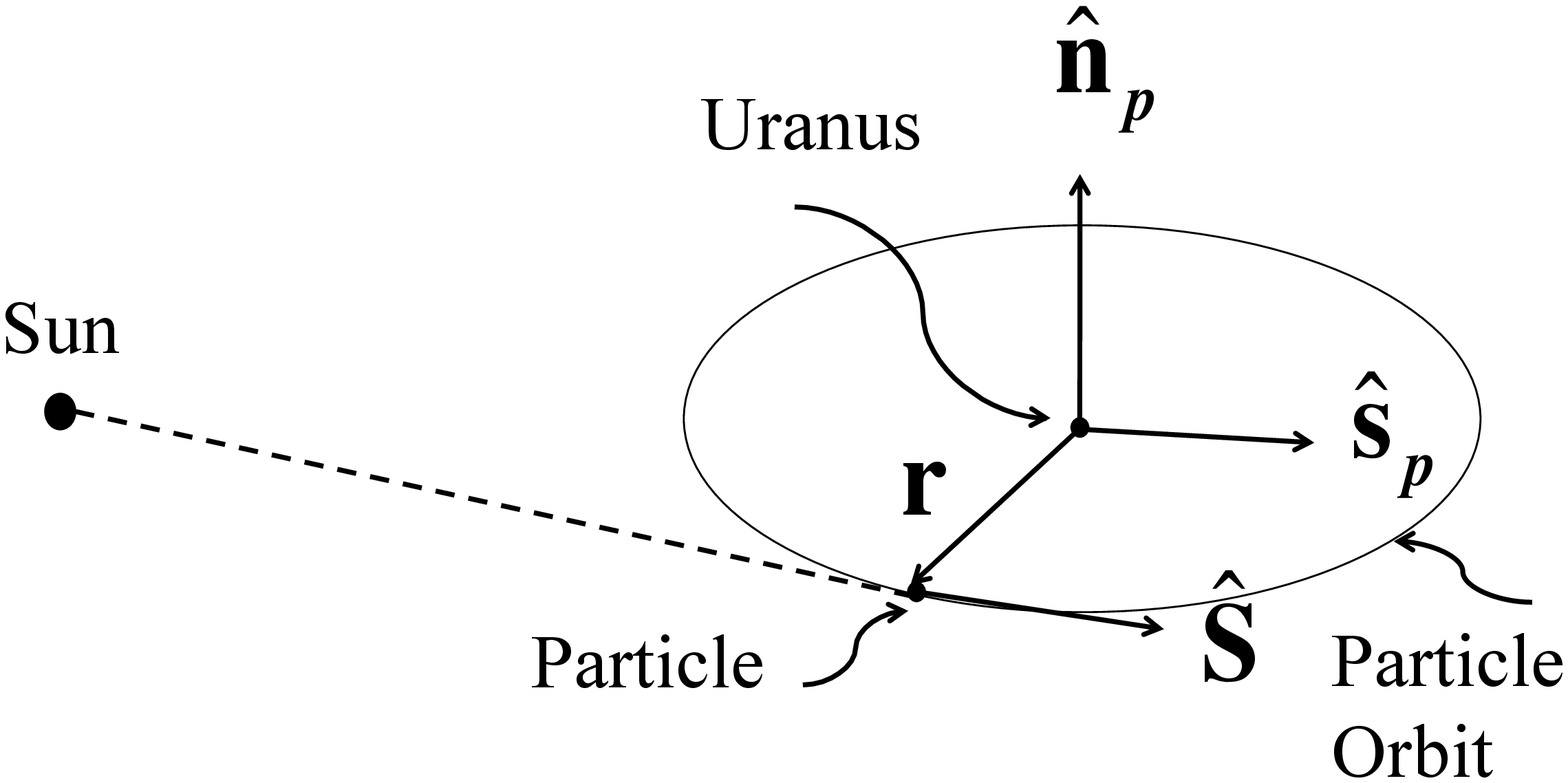}
\caption{\label{vectors}  Diagram showing the various vectors in Eq. \ref{eom}.  The unit vectors ${\bf \hat{n}_p}$ and ${\bf \hat{s}_p}$ point along Uranus' orbit normal and spin axis, respectively.  The unit vector ${\bf \hat{S}}$ points along the line from the Sun to the particle, and the vector ${\bf r}$ is the particle's displacement vector from Uranus.}
\end{figure}

\cite{Tamayo13} investigate the orbital modifications created by this additional force.  The first important consequence of radiation pressure is that it induces a variation in the particle's orbital eccentricity and pericenter location on Uranus' orbital timescale of $\sim 100$ years \citep{Burns79}.  We are interested in much longer secular timescales ($\tau_\text{Sec} \sim 10^4$ yrs), so these fast oscillations can be averaged out.  However, radiation-pressure effects are stronger for smaller particles, leading to larger-amplitude eccentricity variations.  Below a threshold particle size, the eccentricity reaches unity within the first half of a Uranian year, and the grain either collides with the planet or escapes the system.  Integrations show that this threshold size is $\sim 5\mu$m; we therefore only consider particles of radius $10\mu$m and larger.  \cite{Tamayo13} also find that radiation pressure shifts the position of the chaotic zone for most orbits with low initial eccentricities and inclinations to the planet's orbital plane.  While these analytic results furnish good intuition, the Uranian irregular satellites (and therefore the dust grains they generate) lie on high-eccentricity and high-inclination orbits, forcing our detailed investigation to be primarily numerical.  The equation of motion we integrate in our simulations is
\begin{equation} \label{eom}
{\bf \ddot{r}} = - \frac{GM_p}{r^3}{\bf r} + \frac{SAQ_{pr}}{mc} {\bf \hat{S}} - \frac{SA}{mc^2} Q_{pr}[({\bf \dot{r} \cdot \hat{S})\hat{S} + \dot{r}}] - \frac{GM_\odot}{2a_p^3} \nabla[r^2 P_2 ({\bf \hat{n}_p \cdot \hat{r}})] + GM_pR_p^2 J_2' \nabla\Big(\frac{P_2({\bf \hat{s}_p \cdot \hat{r}})}{r^3}\Big),
\end{equation}
where overdots denote time derivatives, and the right-hand terms, in sequence, are due to the dominant Uranian gravity, solar radiation pressure, Poynting-Robertson drag, the Sun's tidal gravity, and Uranus' effective $J_2$, treating the inner satellites' averaged gravity as a contribution to the planet's quadrupole field.  $G$ is the gravitational constant, $M_p$ the planetary mass, $r$ the dust particle's distance from Uranus, $S$ the solar flux at the grain's position, $A$ the particle's cross-sectional area, $Q_{pr}$ the grain's radiation pressure efficiency factor, $m$ the particle mass, $c$ the speed of light, $a_p$ the semi-major axis of Uranus (assumed to be on a circular orbit about the Sun) and $P_2$ the second Legendre polynomial.  The remaining vectors can be seen in Fig. \ref{vectors};  ${\bf r}$ is the particle's displacement vector from Uranus, ${\bf \hat{S}} $ is the unit vector from the Sun to the particle position, ${\bf \hat{n}_p}$ is the unit vector along the planet's orbit normal, and ${\bf \hat{s}_p}$ is the unit vector along Uranus' spin axis.  The effective $J_2$ including the contribution from the inner satellites is denoted by $J_2'$, and is given by
\begin{equation}
J_2' = J_2 + \frac{1}{2} \displaystyle\sum_{i=1}^5 \Big(\frac{m_i}{M_p}\Big) a_i ,
\end{equation}
where $J_2$ is the planetary $J_2$ coefficient, and $m_i$ and $a_i$ are the ith satellite's mass and semimajor axis, respectively.

In order to demonstrate the viability of the creation of the Uranian color dichotomies through dust infall from the irregular satellites, we aim to show four properties of the process:  (i)  the dynamics are capable of spreading dust across the four outermost satellites, rather than concentrating grains on the outermost moon Oberon, (ii) the innermost regular satellite Miranda is not exposed to a comparable effect, (iii) there is an increasing trend with target satellite semimajor axis, and (iv) the incoming dust-particle distribution creates hemispherical leading/trailing asymmetries.  The last point is certainly not obvious in view of Uranus' peculiar obliquity, coupled with the fact that the chaotic orbital evolution of the dust particles generates large-amplitude ($\gtrsim 90^{\circ}$) swings in the grains' orbital inclinations, allowing them to strike the regular satellites from any direction.  The distribution of impacting grains over the regular satellite surfaces is therefore {\it a priori} highly uncertain.

\section{Methods} \label{inicond}
To characterize the dust-transfer efficiency from the irregulars to Uranus' regular satellites (Miranda, Ariel, Umbriel, Titania and Oberon), one would ideally integrate a representative sample of dust particles along with the regular satellites, checking for collisions at each timestep.  Unfortunately, the orbital phase space occupied by the irregulars (and the dust particles they create) is enormous.  One can reduce the computational load by observing that the satellites' and dust particles' pericenters and their relative nodes circulate rapidly and are roughly uniformly distributed from $0$ to $2\pi$.  We verified this for our integrations, though our simulations approximate the motion of the regular satellites as a contribution to the planetary quadrupole \citep{Tamayo13}, and therefore ignore the effects of orbital resonances (we note that orbits almost exclusively destabilize prior to encountering the first-order resonance regions with the regular satellites).  After integrating a dust particle orbit, at each timestep we use the formalism of \cite{Greenberg82}, with corrections from \cite{Bottke93}, to calculate collision probabilities with each of the regular satellites.  These formulae calculate a weighted collision probability assuming the pericenters and relative node to be uniformly distributed.  The values calculated from a single integration are thus equivalent to an average over a much larger population of orbits so distributed.  Finally, we use the methods \cite{Tamayo11} applied in the Saturnian system to combine the probabilities at each timestep into aggregate values.  This approach renders the phase space of initial conditions more tractable.  We believe that the errors introduced by this approximation are not significant relative to the uncertainty in the dust grains' physical and orbital parameters.  In addition to calculating collision probabilities, we further use the techniques in \cite{Tamayo11} to calculate the ratio of material striking the leading vs. the trailing side of each satellite.  We performed all our integrations with the well-established dust integrator dI \citep{Hamilton93} using a Bulirsh-Stoer time-stepper with adaptive stepsize.

Most dust particles escaping irregular-satellite surfaces should have ejection speeds comparable to the satellites' escape velocities \citep{Farinella93}.  Since these escape speeds are much smaller than the satellites' orbital velocities, dust grains should inherit their parent satellites' orbital elements.  While our method outlined above circumvents having to sample a variety of pericenter and node positions, the irregular satellites also span a wide range in semimajor axis ($a$), eccentricity ($e$) and inclination ($i$) \citep{Brozovic09}.  In this paper, we limit ourselves to studying the effects of varying a single initial condition with all other parameters fixed.  For reference initial conditions, we took the approximate mean values across the known irregular satellites:  $a_0 = 460 R_p$, $e_0 = 0.35$ and $i_0 = 155^{\circ}$---where in averaging we have excluded the single prograde irregular, Margaret, and the inclination is measured relative to the Uranian orbital plane \citep{Brozovic09}.  These values should not be taken as necessarily representative of the irregular satellite population over time, as collisions and gravitational perturbations would alter the distribution; rather, they were chosen as sensible (albeit somewhat arbitrary) values for comparison.  We chose a reference particle radius $s$ of $50 \mu$m, which is large enough to be only moderately affected by radiation pressure, and adopted a particle density equal to that of Saturn's irregular satellite Phoebe, the only irregular satellite for which a density has been directly determined ($\rho = 1.6 \text{g cm}^{-3}$, \citealt{Porco05}).  

Taking the above parameters, we performed three suites of integrations, each of which varied a separate initial value.  The first sampled the observed range of irregular-satellite orbital eccentricities ($e_0  = $0.05, 0.2, 0.35, 0.5, 0.65).  The second spanned a range of inclinations $i_0$ ($5^{\circ}, 15^{\circ}, 25^{\circ}, 35^{\circ}, 145^{\circ}, 155^{\circ}, 165^{\circ}, 175^{\circ}$), and the last varied $s$ (10, 20, 30, 50, and $100 \mu$m).  We ignore the intermediate inclinations between $\approx 40-140^{\circ}$ that will undergo Kozai oscillations since the importance of their contribution is unclear, and our methods cannot adequately handle them.  The Kozai effect generates large-amplitude oscillations in the orbital eccentricity.  If its eccentricity amplitude is high-enough, an irregular satellite will collide with a regular satellite on a short timescale, preventing it from generating dust with similar orbital elements.  It is therefore unclear how much these satellites would contribute to the total dust budget.  The reason our methods cannot handle such orbits is that our assumption that the pericenter orientation is uniformly distributed between $[0,2\pi]$ becomes poor.  Beyond a critical inclination that depends on the initial eccentricity, new solutions appear where the pericenter orientation oscillates around a fixed value (librating solutions), whereas circulating solutions cycle at significantly non-uniform rates.  Nevertheless, as discussed in the next section, we expect the behavior of particles undergoing circulating Kozai cycles to match our high-eccentricity cases.

We then calculated the collision probabilities and leading/trailing ratios every 450 yrs.  This choice of timestep carefully samples the secular evolution ($\tau_\text{Sec} \sim 10^4 \text{yrs}$) and changing collision probabilities ($\tau_\text{Col} \sim 10^5 \text{yrs}$), though only captures the fast radiation-pressure induced evolution ($\tau_\text{RP} \sim 100$yrs) in an average sense.  

We started all particles with values of zero for the longitude of ascending node (measured from Uranus' vernal equinox), argument of pericenter, and true anomaly.  While it is true that within a single integration the pericenter and relative node (with any given satellite) are roughly uniformly distributed, the chaotic dynamics in the unstable region mean that slightly different initial choices for these values will yield divergent evolution through the chaotic region.  Performing the collision probability calculations on orbital histories of initially nearby orbits would then yield different (but equally valid) results.  To try and capture this chaotic effect statistically, we instead varied the initial position of the Sun, which also determines the amplitude of the eccentricity oscillation induced by radiation pressure on the fast Uranian orbital timescale \citep{Burns79, Tamayo13}.  Thus, in addition to creating divergent orbital evolution through the chaotic region, this angle choice affects which inner moons can be reached at a given time by changing the amplitude of the fast eccentricity oscillation that is superimposed on the secular evolution.  For each of the eighteen combinations of orbital elements given above (five varying $e$, eight varying $i$ and five varying $s$), we ran integrations for sixteen equally-spaced initial solar positions, yielding a total of two hundred and eighty-eight orbital integrations.  

\section{Results}
For each group of simulations, we plot the ratio of particles striking the leading vs. trailing hemispheres of each satellite, as well as the ``intrinsic" collision probability, defined as the probability of striking a particular moon divided by its total surface area.  To see why this might be a better indicator for the generation of hemispherical asymmetries than a total collision probability, imagine that the infalling dust had an equal probability of striking two moons.  The two satellites would therefore receive equal quantities of dust; however, if one moon were larger than the other, one would expect a larger effect on the smaller satellite since each area element on its surface is subject to a greater quantity of dust.  Normalizing the collision probabilities by the satellite surface areas therefore provides a better comparison.

The results can be qualitatively understood as arising from a competition between three different timescales.  The longest is the P-R timescale over which the semimajor axis (and therefore the pericenter distance) slowly decays.  In the geometric optics limit, this timescale varies linearly with particle size \citep{Burns79}, and for the Uranian system is $\tau_{\text{P-R}} \sim s \times 10^6$ yr, with $s$ in microns.  Next is the much faster secular timescale on which the eccentricity changes when the instablility is reached, $\tau_\text{Sec} \sim 10^4 \text{yrs}$.  Finally, $\tau_{\text{Col}}$ is the typical timescale on which a dust grain collides with a regular satellite once the particle's orbital pericenter dips below the moon's orbital radius, allowing collisions.  While the latter collision time depends on the particle's orbital elements, our numerical calculations show that $\tau_{\text{Col}} \sim 10^5 \text{yrs}$, intermediate between $\tau_\text{Sec}$ and $\tau_{\text{P-R}}$.  

\begin{figure}[!ht]
\includegraphics[width=12cm]{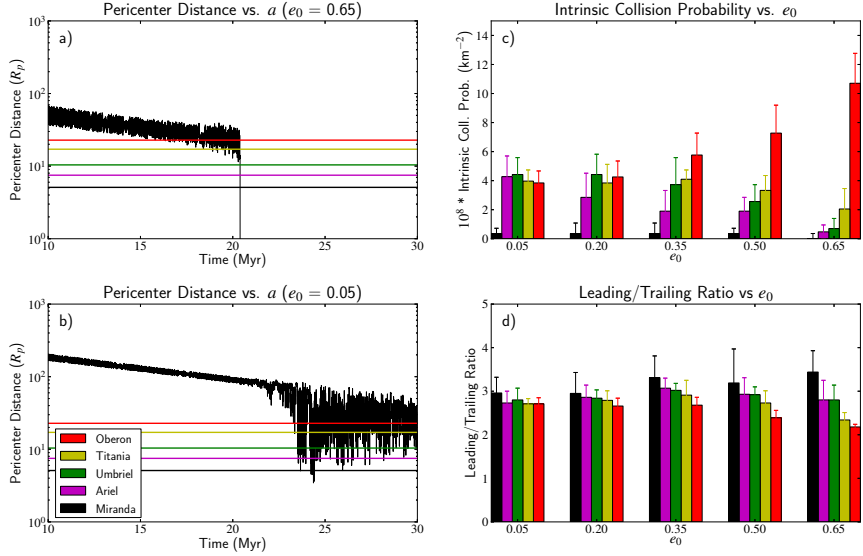}
\caption{\label{e}  Integrations varying $e_0$ (0.05, 0.2, 0.35, 0.5, 0.65) with $a_0 = 460 R_p$, $i_0 = 155^{\circ}$, and $s = 50\mu$m.  Panels a) and b) plot the pericenter distance vs. time for a typical orbit with $e_0 = 0.65$ and $e_0 = 0.05$, respectively.  The semimajor axes of the five regular satellites are plotted as horizontal lines, see the color legend on the figure's bottom left (used in all panels).  Panel c) shows the intrinsic collision probabilities with each target (total collision probability divided by satellite surface area) for the five values of $e_0$.  The bars over each value of $e_0$ are offset and arranged from left to right in order of increasing distance from Uranus.  Thus, Miranda (left) is in black, and Oberon (right) is in red.  Each of the five sets of bar graphs represents an average over sixteen equally spaced initial conditions for the solar position, and the error bars correspond to the standard deviation across those sixteen integrations.  Panel d) displays the ratio of material striking the leading vs. the trailing side of each satellite.}
\end{figure}

Figures \ref{e}a) and \ref{e}b) show two representative cases.  In \ref{e}a), the unstable zone is reached at $t \approx 20.5$ Myr, at which point the orbit's pericenter abruptly plunges inward and the particle strikes Uranus.  However, prior to that at $\sim 15$ Myr, the pericenter occasionally crosses the orbit of the outermost regular satellite Oberon, due to the large orbital eccentricities ($e_0$ = 0.65, which radiation pressure periodically drives even higher).  Then, the pericenter slowly drifts past Oberon on a timescale $\tau_{\text{P-R}} \gg \tau_{\text{Col}}$, allowing that satellite to sweep up most dust particles before they can reach the next moon Titania.  By contrast, in panel \ref{e}b), $e_0$ is low (0.05), and when the orbit reaches the unstable zone at $t \approx 24 Myr$, the pericenter still lies beyond Oberon's orbit.  On the short timescale $\tau_{\text{Sec}}$, the pericenter then plunges inside the orbit of several regular satellites, leading to a more equitable distribution of collision probabilities among the inner moons.  

We expect that particles undergoing circulating Kozai cycles (not integrated), should roughly match the high ($e = 0.65$) case, concentrating most material on Oberon.  Kozai oscillations will periodically drive the eccentricity to high values, allowing the pericenter to cross Oberon's orbit prior to the semimajor axis reaching the unstable range, like in the $e=0.65$ case.  Librating solutions reach lower maximum eccentricities, so this case is more complicated.  It is unfortunately not clear what fraction of irregular satellites would be captured onto librating vs. circulating Kozai trajectories, or how significant the Kozai population is to the total budget of dust generated by the irregular satellites.  If the Kozai population is found to be important, one would have to use alternate methods to ours for the estimation of collision probabilities, \citep[e.g.,][]{Vok12}.

While panels \ref{e}a) and \ref{e}b) showed cases on the extremes of our initial eccentricity distribution, panel \ref{e}c) displays the combined results from all integrations varying $e_0$ (see caption description).  The intrinsic collision probability with Oberon (red) increases substantially for high initial eccentricities, due to the effect discussed in the previous paragraph.  At lower eccentricities, the distribution is more equitable due to the interaction of two factors.  On the one hand, all other aspects being equal, one would expect inner satellites to intercept more dust due to their higher orbital speeds.  Higher relative velocities lead to more frequent encounters between the moon and dust particles, providing more chances for collision.  On the other hand, all other aspects are {\it not} equal, due to the ``random" distribution of successive minima in the pericenter distance upon entering the unstable region (panel \ref{e}b).  A smaller fraction of these minima dip low enough to strike an inner moon than an outer one, leading to a fractionation of collision probability.  Finally, we note that orbital pericenters rarely dip low enough to reach Miranda.  As a result, the intrinsic collision probabilities with Miranda are always much smaller than those with the other moons, consistent with the non-detection of a hemispheric color asymmetry on Miranda by \cite{Buratti91}.  

\begin{figure}[!ht]
\includegraphics[width=12cm]{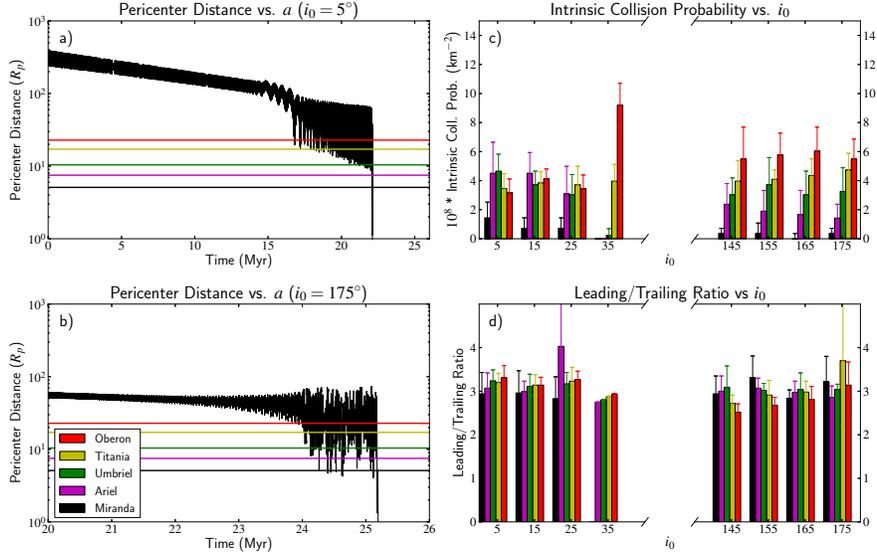}
\caption{\label{i}  Integrations varying $i_0$ ($5^{\circ}, 15^{\circ}, 25^{\circ}, 35^{\circ}, 145^{\circ}, 155^{\circ}, 165^{\circ}, 175^{\circ}$) with $a_0 = 460 R_p$, $e_0 = 0.35$, and $s = 50\mu$m.  Panel a) shows the pericenter distance vs. time for a typical orbit in the prograde group with initial inclination $i_0 = 5^{\circ}$.  Panel b) shows a typical orbit in the retrograde group with $i_0 = 175^{\circ}$.  For an explanation of panels c) and d), see the caption to Fig. \ref{e}.  }
\end{figure}

Panel \ref{e}d) plots the ratio of material striking the leading vs.\ trailing side of each moon.  With Uranus' equatorial plane nearly perpendicular to the ecliptic, it is not {\it a priori} clear what leading/trailing ratio one would expect, particularly since the instability also scatters the orbital inclination over a wide range.  The plotted values are averaged over the sixteen initial solar positions, weighted by their corresponding collision probabilities.  Thus, the leading/trailing ratio in an integration where Miranda receives 0.03$\%$ of dust matters proportionately less than that in a simulation where the collision probability is $3\%$.  The standard deviations are similarly weighted.  The calculated leading/trailing ratios range between $\approx 2-4$.  Our numerical calculations show a definite preference for material striking the leading sides of all the regular satellites.  This is mostly due to the fundamental asymmetry induced by the satellite's motion discussed in the introduction for IDPs and ISDPs.  It is analogous to the increase in rain striking a car's windshield at faster speeds through a storm.  The slight decrease in the leading/trailing ratio with satellite distance from Uranus seems to be a peculiarity of the chosen initial inclination ($i_0 = 155^{\circ}$), as revealed by our next set of integrations (Fig. \ref{i}d).  

Figure \ref{i} shows our results upon varying the initial inclination.  Ignoring for a moment the curious behavior for $i_0 = 35^{\circ}$ in panel \ref{i}c), the collision probabilities do not seem to depend strongly on $i_0$, though systematic differences occur between prograde and retrograde orbits.  The distribution is more equitable for prograde particles, while for retrograde particles the intrinsic collision probabilities increase with the target's distance from Uranus.  This is due to a difference in the manner the chaotic region is approached, visible in panels \ref{i}a) and \ref{i}b).  For prograde particles (panel a), the orbital eccentricity spikes abruptly upon entering the unstable region, and the pericenter plunges.  The dust can thereby access all the moons on a single $\tau_\text{Sec}$, leading to a flatter probability distribution among the satellites.  By contrast, the eccentricities of retrograde particles undergo gradual growth as the chaotic regime is approached.  The pericenter distance is thereby often able to dip inside Oberon's orbit (panel \ref{i}b) prior to being exposed to the remainder of the moons, concentrating the collision probability on the outermost satellite.  We note that each individual excursion inside Oberon's orbit is significant since $\tau_\text{Col}$ is not much longer than $\tau_\text{Sec}$, the relevant timescale on which the eccentricity evolves.

The dynamical cause of this prograde/retrograde asymmetry is not clear.  Similarly curious is the high collisional likelihood on Oberon for $i_0 = 35^{\circ}$.  We find upon inspection of our integrations that at such high inclinations below, but approaching, the critical $i_0$ at which Kozai oscillations occur ($i_0 = 43.5^{\circ}$ for $e_0 = 0.35$), the system undergoes nearly regular Kozai-like oscillations prior to reaching the chaotic semimajor axis regime.  These high eccentricities bring the pericenter inside Oberon's orbit, but outside the orbital radii of the remaining moons, thus concentrating material on Oberon.  The same does not occur for highly inclined retrograde orbits.  An analysis of the dynamics in this rich and highly non-linear regime is beyond the scope of this paper.  Low-eccentricity orbits close to the equilibrium Laplace plane are treated by \cite{Tamayo13}.  

As in the case varying $e_0$, the leading/trailing ratio $\approx 3$, independent of $i_0$ (panel d of Fig. \ref{i}).  There is no systematic trend with target distance from Uranus.  

\begin{figure}[!ht]
\includegraphics[width=12cm]{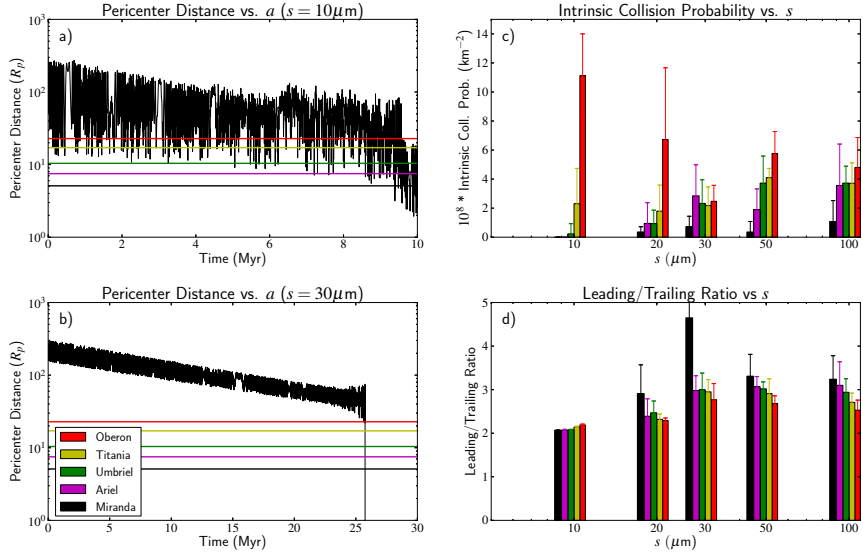}
\caption{\label{s}   Integrations varying $s$ (0, 20, 30, 50, and $100 \mu$m) with $a_0 = 460 R_p$, $e_0 = 0.35$, and $i_0 = 155^{\circ}$.  Panel a) shows the pericenter distance history vs. time for a typical $10 \mu$m particle orbit.  Panel b) shows a typical $30 \mu$m grain orbit.  For an explanation of panels c) and d), see the caption to Fig. \ref{e}.}
\end{figure}

Figure \ref{s} shows our results for various grain sizes.  Smaller particles, more affected by radiation pressure, have larger-amplitude eccentricity oscillations ($\tau_\text{RP} \sim 100$ yrs) superimposed on their secular evolution.  Radiation pressure also shifts the position at which orbits become unstable \citep{Tamayo13}.  Panel c) shows that the collision probability distribution for 50 and $100\mu$m grains are qualitatively similar.  Due to the enhanced effects of radiation pressure, we find that the orbital eccentricities of 20 and $30\mu$m particles reach unity within one to a few $\tau_\text{Sec}$ of entering the unstable region (panel b), and they strike Uranus.  The available time for impact with satellites is thus drastically reduced.  While in all of the previously discussed cases the fraction of particles striking Uranus was $\lesssim 5\%$, approximately $30\%$ and $40\%$ of particles strike the planet in the $20 \mu$m and $30\mu$m cases, respectively.  For 10$\mu$m particles, and for some 20$\mu$m grains, the rapid eccentricity oscillations induced by radiation pressure ($\tau_\text{RP} \sim 100$ yrs) are of such large amplitude that for some initial solar positions the pericenter dips inside Oberon's orbit before the semimajor axis reaches the unstable region (see panel a).  This concentrates the probability distribution on the outermost satellite.  We note that, especially for the 10$\mu$m particles, the orbital behavior varies substantially across different initial solar positions.  The above statement nevertheless remains qualitatively valid.  

From panel d), one can see that the leading/trailing ratio remains $\approx 3$ for $s \gtrsim 30 \mu$m.  While we cannot simply explain the decreased asymmetry for smaller particles, it remains true that dust grains will preferentially strike the leading sides of the regular satellites.  We performed analogous integrations for prograde particles ($i_0 = 25^{\circ}$, not plotted) of varying sizes, and found the distributions to be qualitatively similar in all the respects discussed above, with the exception that the anomalously high leading/trailing ratio on Miranda for $30 \mu$m particles disappeared, and similar spikes occurred for Titania at $s=20\mu$m and Ariel at $s=50\mu$m.  In all cases, the standard deviations are large, so we do not believe them to be observationally important.

\section{Conclusion}
In this paper we have tried to find a mechanism capable of explaining the hemispherical asymmetries on the Uranian regular satellites found by \cite{Buratti91}.  We argued that various sources beyond the Uranian system and magnetospheric effects are either incapable of producing leading/trailing differences, or in cases where they might create asymmetries, that the effect should be largest for the inner satellites.  This predicted trend is opposite to what is observed.  Innermost Miranda has no detectable asymmetry, and hemispherical differences grow with semimajor axis for the outer four moons \citep{Buratti91}.  

We then investigated the dynamics of infalling dust from the irregular satellites as a possible mechanism.  This process is complicated by the fact that Uranus' extreme obliquity causes chaotic large-amplitude variations in particles' orbital eccentricity and inclination over a range of semimajor axes \citep{Tremaine09, Tamayo13}.  We found that:  (i) dust reaches the outermost four satellites, contrary to the expectation (in the absence of this dynamical instability) that the vast majority of dust would be concentrated on the outermost moon Oberon, (ii) dust-orbit pericenters rarely reach the semimajor axis of Miranda, consistent with the fact that this innermost moon shows no hemispherical asymmetry (cf. panel c of Figs. \ref{e}-\ref{s}), (iii) for retrograde dust particles, the intrinsic collision probability with each satellite tends to increase with semimajor axis (panel c of Figs. \ref{e}-\ref{s}), and (iv) despite the unusual Uranian geometry and the fact that the mentioned instability leads to chaotically varying orbital inclinations, approximately three times more dust strikes the leading hemispheres of each of the regular satellites than their respective trailing hemispheres, independent of initial conditions ($\approx 2$ times for the smallest dust grains); see panel d) of Figs. \ref{e}-\ref{s}.  

We point out that not all initial conditions generate collision probabilities that increase with the semimajor axis of the target satellite.  Most notably, only retrograde particles with moderate-to-high orbital eccentricity do.  The directionality fits well with the fact that retrograde irregular satellites are more dynamically stable than prograde moons and are therefore longer-lived \citep{Carruba02, Nesvorny07}.  Today, out of nine known Uranian irregulars, only Margaret is prograde.  The eccentricity requirement does not preclude a population of low-eccentricity impactors (that would tend to produce comparable color asymmetries across the satellites)---it only requires that a significant fraction of the dust population be born on moderate-to-high eccentricity orbits to produce the trend.  If the current irregular satellite orbital eccentricity distribution is any indicator of this historical average, this seems plausible.

We therefore conclude that, despite the uncertainties, infall of dust from the irregular satellites furnishes the best explanation for the color asymmetries on the Uranian regular satellites.  If this hypothesis is correct, it implies that the lifetimes of dust grains orbiting Uranus at large semimajor axes ($\sim 100 R_p$) are much longer ($\gtrsim 10$ Myr) than in typical planetary magnetospheres.  It would also require that irregular-satellite dust infall overwhelm the flux from sources that would generate the opposite trend with semimajor axis, in particular, IDPs.  The flux of IDPs over the Solar System's history is quite uncertain, but one can obtain a rough lower limit for the total interplanetary dust mass intercepted by the Uranian satellites by extrapolating the current flux backward in time.  Using estimates for the current flux of IDPs in the outer Solar System by \cite{Cuzzi98}, this demands a dust mass generated by the irregulars $\gtrsim 10^{14}$kg over the system's history.  While no current dust ring associated with the irregular satellites has been detected thus far around Uranus, the production of dust through mutual collisions between irregulars should have been been strongly concentrated in the first few hundred Myr after the capture of the irregular satellites \citep{Bottke10}.  An analogous diffuse dust ring has been discovered around Saturn with the Spizer Space Telescope; it is generated by the irregular satellite Phoebe \citep{Verbiscer09}.  This Phoebe ring has an estimated mass $\sim 3$ orders of magnitude smaller than our lower limit of $10^{14}$ kg.  Thus, if our hypothesis is correct, the irregular satellites must have generated a substantial amount of dust in the past.

\section{Acknowledgements}
We thank Philip Nicholson, Matthew Hedman, Matthew Tiscareno, and Rebecca Harbison for their insightful comments.  This work was supported by the Cassini project and NASA's Planetary Geology and Geophysics Program.


\newcommand{\icarus}{Icarus}
\newcommand{\aj}{AJ}
\newcommand{\araa}{ARA\&A}
\newcommand{\apj}{ApJ}
\newcommand{\apjl}{ApJ}
\newcommand{\apjs}{ApJS}
\newcommand{\ao}{Appl.~Opt.}
\newcommand{\apss}{Ap\&SS}
\newcommand{\aap}{A\&A}
\newcommand{\aapr}{A\&A~Rev.}
\newcommand{\aaps}{A\&AS}
\newcommand{\azh}{AZh}
\newcommand{\baas}{BAAS}
\newcommand{\jrasc}{JRASC}
\newcommand{\memras}{MmRAS}
\newcommand{\mnras}{MNRAS}
\newcommand{\pra}{Phys.~Rev.~A}
\newcommand{\prb}{Phys.~Rev.~B}
\newcommand{\prc}{Phys.~Rev.~C}
\newcommand{\prd}{Phys.~Rev.~D}
\newcommand{\pre}{Phys.~Rev.~E}
\newcommand{\prl}{Phys.~Rev.~Lett.}
\newcommand{\pasp}{PASP}
\newcommand{\pasj}{PASJ}
\newcommand{\qjras}{QJRAS}
\newcommand{\skytel}{S\&T}
\newcommand{\solphys}{Sol.~Phys.}
\newcommand{\sovast}{Soviet~Ast.}
\newcommand{\ssr}{Space~Sci.~Rev.}
\newcommand{\zap}{ZAp}
\newcommand{\nat}{Nature}
\newcommand{\iaucirc}{IAU~Circ.}
\newcommand{\aplett}{Astrophys.~Lett.}
\newcommand{\apspr}{Astrophys.~Space~Phys.~Res.}
\newcommand{\bain}{Bull.~Astron.~Inst.~Netherlands}
\newcommand{\fcp}{Fund.~Cosmic~Phys.}
\newcommand{\gca}{Geochim.~Cosmochim.~Acta}
\newcommand{\grl}{Geophys.~Res.~Lett.}
\newcommand{\jcp}{J.~Chem.~Phys.}
\newcommand{\jgr}{J.~Geophys.~Res.}
\newcommand{\jqsrt}{J.~Quant.~Spec.~Radiat.~Transf.}
\newcommand{\memsai}{Mem.~Soc.~Astron.~Italiana}
\newcommand{\nphysa}{Nucl.~Phys.~A}
\newcommand{\physrep}{Phys.~Rep.}
\newcommand{\physscr}{Phys.~Scr}
\newcommand{\planss}{Planet.~Space~Sci.}
\newcommand{\procspie}{Proc.~SPIE}

\end{document}